\documentclass[10pt,twocolumn,english]{revtex4}
\usepackage[T1]{fontenc}
\usepackage[latin9]{inputenc}
\usepackage[a4paper]{geometry}
\usepackage{amsmath}
\usepackage{amssymb}
\usepackage{graphicx}



\usepackage{babel}

\begin{document}

\title{Renormalization Group Flow of Hexatic Membranes}

\author{Alessandro Codello$^a$ and Omar Zanusso$^b$\\
$^a$\emph{SISSA, via Bonomea 265, I-34136 Trieste, Italy}\\
$^b$\emph{Radboud University, Nijmegen, Institute for Mathematics, Astrophysics and Particle Physics, Heyendaalseweg 135, 6525 AJ Nijmegen, The Netherlands}
}

\begin{abstract}
We investigate hexatic membranes embedded in Euclidean $D$-dimensional
space using a re--parametrization invariant formulation combined with
exact renormalization group (RG) equations. An \emph{XY}--model coupled
to a fluid membrane, when integrated out, induces long--range interactions
between curvatures described by a Polyakov term in the effective action.
We evaluate the contributions of this term to the running surface tension, bending
and Gaussian rigidities in the approximation of vanishing disclination (vortex) fugacity.
We find a non--Gaussian fixed--point where the membrane is crinkled and  has a non--trivial fractal dimension.
\end{abstract}

\maketitle

\subsection{Introduction}

Hexatic membranes are a very interesting example of physical system
which undergoes a non--trivial phase transition induced by a combination
of geometrical and topological interactions \cite{Nelson_Piran_Weinberg_1988}. 
This system is even more interesting since it is characterized by local re--parametrization invariance,
and is thus an example of a theory characterized by local symmetries with a
renormalization group (RG) flow exhibiting a non--Gaussian fixed point  in two dimensions.

In the continuum, Hexatic membranes are described as a fluid membrane coupled to an $O(2)$--symmetric scalar field.
The $O(2)$--global symmetry represents in the continuum the $\mathbb{Z}_6$--symmetry characteristic
of the orientational order of triangular or hexagonal lattices (from this last it takes the name) that melt \cite{Nelson_Piran_Weinberg_1988};
the $O(2)$--symmetric scalar field plays thus the role of the orientational order parameter. 
Hexatic membranes are effectively described by combining the fluid membrane and the $O(2)$--model actions:
\begin{eqnarray}
S[\mathbf{N},\mathbf{r}]=\int d^{2}x\sqrt{g}\left\{ \mu+\frac{\kappa}{2}K^{2}+\frac{\bar{\kappa}}{2}R\right\}\qquad\nonumber \\
+\frac{K_{A}}{2}\int d^{2}x\sqrt{g}\,\nabla_{\alpha}\mathbf{N}\cdot\nabla^{\alpha}\mathbf{N}\,.\label{1}
\end{eqnarray}
In (\ref{1}) $K_{A}$ is the hexatic stiffness, $\mu$ is the surface tension, $\kappa$ is the bending rigidity and $\bar{\kappa}$ is the Gaussian
rigidity \cite{Nelson_Piran_Weinberg_1988}.
The action (\ref{1}) is invariant under re--parametrization of the embedding field $\mathbf{r}:\mathbb{R}^2\rightarrow\mathbb{R}^D$;
$g_{\mu\nu}=\partial_\mu\mathbf{r}\cdot\partial_\nu\mathbf{r}$ is the induced metric, $K^2$ is the square of the trace of the extrinsic curvature, while $R$ is the intrinsic Gaussian curvature.
The scalar field, tangent to the surface, is constrained to have unit length $\mathbf{N}\cdot\mathbf{N}=1$.

In this paper we study the RG flow of the couplings appearing in the action (\ref{1})
by computing their beta functions for arbitrary embedding dimension $D$.
Differently from previous approaches to hexatic membranes \cite{Nelson_Peliti_1987,David_Guitter_Peliti_1987,Guitter},
we use a geometrical formulation of the model which is explicitly covariant 
and we apply to it the effective average action formalism \cite{Berges:2000ew},
which has been adapted to fluid membranes in \cite{Codello_Zanusso_2011}.
The phase diagram of fluid membranes
is modified by the inclusion of the hexatic stiffness,
which induces, in the RG flow, a non--Gaussian IR fixed--point at which the surface is \emph{crinkled},
i.e. it has a non--trivial finite fractal dimension \cite{David_Guitter_Peliti_1987}.

The $O(2)$--action in (\ref{1}) can also be seen as the generalization on a membrane of the continuum action describing the \emph{XY}--model.
As it is well know, the \emph{XY}--model is subject to the Kosterlitz--Thouless topological phase transition \cite{Kosterlitz}, mediated by the unbinding of vortices.
In the context of hexatic membranes, disclinations \cite{Nelson_Piran_Weinberg_1988} are the relevant topological excitations;
these interact by long--range Coulomb interactions, and
in this way they affect the phase behavior of the model.
Interestingly, also the Gaussian curvature plays a similar role, since it is a source of frustration for the field $\mathbf{N}$,
which, when parallel transported around a closed loop, is shifted proportionally to the total Gaussian curvature enclosed.
This is described by long--range Coulomb interactions between Gaussian curvatures in the form of a Polyakov term $\int\sqrt{g}R\frac{1}{\Delta}R$ \cite{Polyakov:1981rd},
resulting from the integration of the $O(2)$--field.

Even if, in the effective average action formalism, it is possible to account for the effect of topological defects
without explicitly summing over them \cite{Grater:1994qx},
in this paper we will discard the contribution of disclinations and focus on the effect that the
long--range Coulomb interactions between Gaussian curvatures have on the phase diagram.
%
Our main interest is to understand the effect that non--local invariants
have on the RG flow of the couplings of the local ones.
%
As we will see, it is the hexatic stiffness $K_A$, seen as the coupling of the Polyakov term,
that has the effect of changing the phase structure of fluid membranes by modifying
the beta functions of the bending rigidity.

\subsection{Induced long-range curvature--curvature interactions}

At every point of the membrane we can introduce zwei-bein orthonormal vector fields $\mathbf{e}_a=e^{\alpha}_{a}\partial_{\alpha}\mathbf{r}$ such that $e^{\alpha}_{a}e^{\beta}_{b}g_{\alpha\beta}=\delta_{ab}$.
Covariant derivatives of the zwei-beins define the spin connection $\omega_{\alpha ab}$ through the relation $\nabla_{\alpha}\mathbf{e}_a=\omega_{\alpha ab}\mathbf{e}_b$, which describes how the zwei-beins rotates under parallel transport.
We can now write the vector field $\mathbf{N}$ in the zwei-bein basis $\mathbf{N}=\cos \theta \, \mathbf{e}_1+\sin \theta \, \mathbf{e}_2$, defining in this way the angle variable $\theta$, in terms of which the action (\ref{1}) becomes:
\begin{eqnarray}
S[\theta,\mathbf{r}]=\int d^{2}x\sqrt{g}\left\{ \mu+\frac{\kappa}{2}K^{2}+\frac{\bar{\kappa}}{2}R\right\}\qquad\qquad\nonumber \\
+\frac{K_{A}}{2}\int d^{2}x\sqrt{g}\,(\partial_{a}\theta+\omega_{a})(\partial^{a}\theta+\omega^{a})\,.\label{2.1}
\end{eqnarray}
The angle $\theta$ parametrizes the one dimensional sphere $O(2)=S^1$.
Since the field $\theta$ appears quadratically, we can complete the square to perform the Gaussian
integral obtaining in this way an effective action for the membrane alone.
This amounts in shifting the field $\theta\rightarrow\theta-\int \frac{1}{\Delta}\nabla_{a}\omega^{a}$
of the amount given by the solution of the tree--level equation of motion $\Delta\theta+\nabla_{a}\omega^{a}=0$,
where $\Delta\equiv-g^{\alpha\beta}\nabla_{\alpha}\nabla_{\beta}$ is the covariant Laplacian.
After using the relation between the Gaussian curvature and the spin connection, $R=\frac{2}{\sqrt{g}}\epsilon_{ab}\nabla^{a}\omega^{b}$,
one finds that the action (\ref{2.1}) becomes:
\begin{eqnarray}
S[\theta,\mathbf{r}]&&=\int d^{2}x\sqrt{g}\left\{ \mu+\frac{\kappa}{2}K^{2}+\frac{\bar{\kappa}}{2}R\right\}\nonumber\\
&&+\frac{K_{A}}{2}\int d^{2}x\sqrt{g}\,\partial_{a}\theta\partial^{a}\theta\nonumber\\
&&+\frac{K_{A}}{8}\int d^{2}x\sqrt{g}R\frac{1}{\Delta}R\,.\label{2.2}
\end{eqnarray}
The step between (\ref{2.1}) and (\ref{2.2}) is analogous to what happens in Liouville theory,
where the Gaussian action $\int\sqrt{g}\left(\frac{1}{2}\phi\Delta\phi+\phi R\right)$
gives rise to a Polyakov term with both classical and quantum contributions.
In the Liouville case one makes the shift $\phi\rightarrow\phi-\int\frac{1}{\Delta}R$
obtained by solving the tree--level equation of motion $\Delta\phi+\phi R=0$.
The underlining idea is that the tree--level equations of motion directly give the correct change of variables
to complete the square in a Gaussian integral.

The $O(2)$--field has the effect of inducing long--range interactions
between Gaussian curvature; these are represented in (\ref{2.2}) by a term proportional
to the Polyakov effective action.
At this point, as done in \cite{David_Guitter_Peliti_1987}, one can integrate out the field
$\theta$ to obtain the hexatic membrane effective action:
\begin{eqnarray}
S_{eff}[\mathbf{r}]=\int d^{2}x\sqrt{g}\left\{ \mu_{0}+\frac{\kappa_{0}}{2}K^{2}+\frac{\bar{\kappa}_{0}}{2}R\right\} \nonumber\\
+\frac{K_{A,0}}{8}\int d^{2}x\sqrt{g}R\frac{1}{\Delta}R\,,\label{3}
\end{eqnarray}
with renormalized $\mu_0$, $\bar{\kappa}_0$ (the bending rigidity is not renormalized $\kappa_0=\kappa_\infty$) and finitely renormalized hexatic stiffness:
\begin{equation}
K_{A,0}=K_{A,\infty}-\frac{1}{12\pi}\,.\label{KA}
\end{equation}
One can obtain (\ref{3}) using the effective average action formalism along the lines of \cite{Codello_2010},
where it was shown how the Polyakov action arises as field modes are integrated out,
one by one, from the UV to IR, and resulting in (\ref{KA}).

\subsection{Flow equation and beta functions}

The effective average action is a functional that depends on an infrared
scale $k$ and that interpolates smoothly between the bare, or microscopic,
action for $k\rightarrow\infty$ and the full effective action for
$k\rightarrow0$; when local symmetries are present it is constructed
using the background field method. The adaptation of the formalism
to fluid membranes has already been done \cite{Codello_Zanusso_2011} to which we
refer for more details. For a complementary study of polymerized membranes using the effective average action approach see \cite{Kownacki_Mouhanna_2009}.
One can construct a functional $\Gamma_{k}\left[\mathbf{r}\right]$
such that it is a re--parametrization invariant functional of the embedding
field $\mathbf{r}$ of the membrane.

The main advantage of working with the effective average action formalism is that we dispose of
an exact RG flow equation that describes its scale derivative ($t=\log k$):
\begin{equation}
\partial_{t}\Gamma_{k}[\mathbf{r}]=\frac{1}{2}\textrm{Tr}\frac{\partial_{t}R_{k}(\Delta)}{\Gamma_{k}^{(2)}[\mathbf{r}]+R_{k}(\Delta)}\,.\label{2}
\end{equation}
In (\ref{2}) $\Gamma_{k}^{(2)}[\mathbf{r}]$ represents the Hessian
obtained by expanding the effective average as $\mathbf{r}\rightarrow\mathbf{r}+\nu^{i}\mathbf{n}^{i}$
to second order in the fluctuation fields $\nu^{i}$. Here $\nu^{i}$,
with $i=D-2$, are the fluctuations in the direction of the $\mathbf{n}^{i}$ 
normals vectors and the function $R_{k}(\Delta)$ is the cutoff kernel. The derivation
of the flow equation (\ref{2}) is also given \cite{Codello_Zanusso_2011}.

\begin{figure}
\begin{centering}
\includegraphics[scale=0.7]{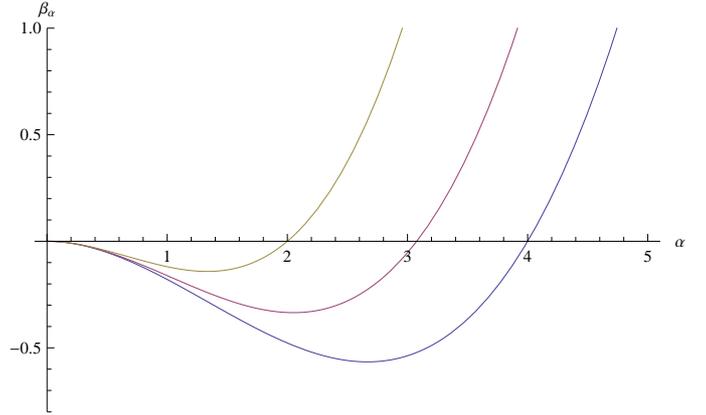}
\par\end{centering}
\caption{Beta function of the coupling $\alpha_{k}$ for (from bottom) $K_{A}=\frac{1}{2},1,2$.}
\end{figure}
%
To proceed we need to make a truncation ansatz for the effective average
action in order to project the RG flow to a treatable finite dimensional
subspace. We choose the scale dependent version of (\ref{3}):
\begin{eqnarray}
\Gamma_{k}[\mathbf{r}]=\int d^{2}x\sqrt{g}\left\{ \mu_{k}+\frac{\kappa_{k}}{2}K^{2}+\frac{\bar{\kappa}_{k}}{2}R\right\}\nonumber\\
+\frac{K_{A,k}}{8}\int d^{2}x\sqrt{g}R\frac{1}{\Delta}R\,,\label{3.1}
\end{eqnarray}
where we introduced the running surface tension $\mu_{k}$, the running
bending $\kappa_{k}$ and Gaussian $\bar{\kappa}_{k}$ rigidities.

The hexatic stiffness $K_{A,k}$ does not renormalize, i.e. its beta function is zero:
\begin{equation}
\partial_t K_{A,k}=0\,. \label{KA0}
\end{equation}
%
One way to see this, following the arguments of \cite{David_Guitter_Peliti_1987},
is to consider the Polyakov term in the effective action (\ref{3}) as the result of the integration of $N$ free scalar fields;
then $N=K_{A,0}=K_{A,\infty}-\frac{1}{12\pi}$ is preserved by fluctuations.
Another way to reach the same conclusion is to consider the $\theta$ scalar field as a non--linear sigma model on $O(2)=S^1$. The beta function of $K_{A,k}$ is then proportional to the Gaussian curvature of $S^1$  \cite{Codello:2008qq}, which is identically zero since the manifold is one--dimensional.
A third argument is related to the flow of the Polyakov action. In \cite{Codello_2010} it is shown that the RG flow induced by a minimally coupled scalar, as is $\theta$, induces a Polyakov term in the effective average action only in the limit $k\rightarrow0$; thus for non--zero $k$, the flow of $K_{A,k}$, as extracted form the Polyakov term, is zero.
In the following we will consider $K_{A,k}=K_A$ as a parameter.
As in the standard Kosterlitz--Thouless topological phase transition, the flow of the hexatic stiffness is induced by topological defects (here the disclinations)  \cite{Park_Lubensky_1996a}.
We will discuss this in a future work \cite{Codello_Mouhanna_Zanusso_2013}.
\begin{figure}
\begin{centering}
\includegraphics[scale=0.5]{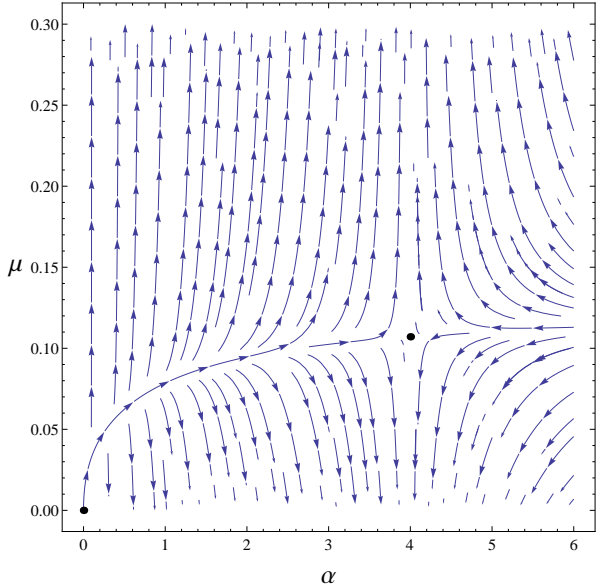}
\par\end{centering}
\caption{The flow in the $(\alpha_{k},\tilde{\mu}_{k})$ plane for $D=3$ and $K_{A}=1$. The Gaussian and non--Gaussian fixed points are marked dots.}
\end{figure}

To obtain the projected flow for the couplings in (\ref{3.1}) we proceed as in
\cite{Codello_Zanusso_2011} with the difference that we need to determine
the contribution of the hexatic stiffness to the running couplings,
i.e. we need to compute the terms proportional to $\int \sqrt{g}$,
$\int \sqrt{g}K^{2}$ and $\int \sqrt{g}R$ on the rhs
of the flow equation (\ref{2}) generated by the Hessian of the Polyakov
term in (\ref{3.1}). We report the details of the computation of
the Hessian in the appendix. Following the notation of \cite{Codello_Zanusso_2011}, we find:
\begin{eqnarray}
\Gamma_{k}^{(2)}[\mathbf{r}]_{ij}&=&\left(\kappa_{k}\Delta^{2}+\mu_{k}\Delta\right)\delta_{ij}\nonumber
+\left(V_{ij}^{\alpha\beta}+W_{ij}^{\alpha\beta}\right)\nabla_{\alpha}\nabla_{\beta}\nonumber\\
&&+\,U_{ij}+O(K^{4})\,,\label{3.01}
\end{eqnarray}
where $V_{ij}^{\alpha\beta}$ and $U_{ij}$ are as in \cite{Codello_Zanusso_2011}
and
\begin{equation}
W_{ij}^{\alpha\beta}=K_{A}\left[\frac{1}{2}\delta_{ij}R+\frac{1}{2}K^{i}K^{j}+K_{\gamma\delta}^{i}K^{j}{}^{\gamma\delta}\right]g^{\alpha\beta}\,,\label{3.2}
\end{equation}
is the contribution to the Hessian coming form the Polyakov action.
In (\ref{3.01}) and (\ref{3.2}) we kept terms up to order $K^{2}$
and $R$. Following the procedure of \cite{Codello_Zanusso_2011}
to perform the functional trace and using the contraction 
%
$
g_{\alpha\beta}W_{ii}^{\alpha\beta}=K_{A}\left[2\left(D-3\right)R+3K^{2}\right]
$,
%
finally gives the beta functions:
\begin{eqnarray}
\partial_{t}\mu_{k} & = & \frac{D-2}{8\pi}Q_{1}\left[G_{k}\partial_{t}R_{k}\right]\nonumber \\
\partial_{t}\kappa_{k} & = & \frac{D}{8\pi}Q_{2}\left[G_{k}^{2}\partial_{t}R_{k}\right]\kappa_{k}-\frac{3}{32\pi}Q_{2}\left[G_{k}^{2}\partial_{t}R_{k}\right]K_{A}\nonumber \\
\partial_{t}\bar{\kappa}_{k} & = & \frac{D-2}{24\pi}Q_{0}\left[G_{k}\partial_{t}R_{k}\right]-\frac{1}{4\pi}Q_{1}\left[G_{k}^{2}\partial_{t}R_{k}\right]\mu_{k}\nonumber \\
 &  & +\frac{D-3}{32\pi}Q_{2}\left[G_{k}^{2}\partial_{t}R_{k}\right]K_{A}\,.\label{6}
\end{eqnarray}
In (\ref{6}) the $Q$--functionals are defined as
\begin{equation}
Q_{n}[h]=\left\{ \begin{array}{ccc}
\frac{1}{\Gamma(n)}\int_{0}^{\infty}dz\, z^{n-1}h(z) &  & n>0\\
(-1)^{n}h^{(n)}(0) &  & n\leq0
\end{array}\right.\,,\label{6.1}
\end{equation}
and we introduced the fourth--order regularized propagator:
\begin{equation}
G_{k}(z)=\frac{1}{\kappa_{k}z^{2}+\mu_{k}z+R_{k}(z)}\,.\label{7}
\end{equation}
The beta functions (\ref{6}) are valid for general embedding dimension
$D$ and for any admissible cutoff shape function $R_{k}(z)$. They
are the main result of this section.

The effect of the inclusion of the hexatic stiffness through the Polyakov
term in (\ref{3.1}) is a modified renormalization of the bending
and Gaussian rigidities.
For $K_{A}=0$ we reproduce the system of
beta functions found in \cite{Codello_Zanusso_2011,Forster_1986}.
Note that the additional renormalization of the Gaussian rigidity
vanishes in the physically relevant case $D=3$.

\subsection{Crinkled phase}


We can start to consider the beta function for the bending and Gaussian
rigidities when $\mu_{k}=0$. Employing the cutoff shape function $R_{k}(z)=(k^{4}-z^2)\theta(k^{2}-z)$
in (\ref{6}) we find:
\begin{eqnarray}
\partial_{t}\kappa_{k} & = & \frac{1}{4\pi}\left(D-\frac{3}{4}\frac{K_{A}}{\kappa_{k}}\right)\nonumber \\
\partial_{t}\bar{\kappa}_{k} & = & \frac{D-8}{6\pi}+\frac{D-3}{16\pi}K_{A}\alpha_{k}\,.\label{8}
\end{eqnarray}
The beta function of $\kappa_{k}$ agrees with \cite{David_Guitter_Peliti_1987},
while the beta function of $\bar{\kappa}_{k}$ is new. In terms of
the couplings $\alpha_{k}=1/\kappa_{k}$ and $\bar{\alpha}_{k}=1/\bar{\kappa}_{k}$
we find:
\begin{eqnarray}
\partial_{t}\alpha_{k} & = & -\frac{\alpha_{k}^{2}}{4\pi}\left(D-\frac{3}{4}K_{A}\alpha_{k}\right)\nonumber \\
\partial_{t}\bar{\alpha}_{k} & = & -\frac{\bar{\alpha}_{k}^{2}}{6\pi}\left(D-8+(D-3)K_{A}\alpha_{k}\right)\,. \label{Baa}
\end{eqnarray}
%
%
We observe that the system (\ref{Baa}) has a non--trivial fixed--point at:
\begin{equation}
\alpha_{*}=\frac{4D}{3}\frac{1}{K_{A}}\qquad\qquad\bar{\alpha}_{*}=0\,,\label{10}
\end{equation}
together with the Gaussian one $\alpha_{*}=\bar\alpha_{*}=0$.
%
Note that $\alpha_{*}$ depends inversely
on $K_{A}$, thus the non--Gaussian fixed--point can be controlled in
a perturbative expansion for large hexatic stiffness.
%

When we re--introduce the surface tension and use the dimensionless
variable $\mu_{k}=k^{2}\tilde{\mu}_{k}$ the flow in the $(\alpha_{k},\tilde{\mu}_{k})$ plane
is described by the following beta functions:
\begin{widetext}
\begin{eqnarray}
\partial_{t}\tilde{\mu}_{k} & = & -2\tilde{\mu}_{k}-\frac{D-2}{2\pi\sqrt{4+\tilde{\mu}_{k}^{2}-\frac{4}{\alpha_{k}}}}\log\frac{2+\tilde{\mu}_{k}-\sqrt{4+\tilde{\mu}_{k}^{2}-\frac{4}{\alpha_{k}}}}{2+\tilde{\mu}_{k}+\sqrt{4+\tilde{\mu}_{k}^{2}-\frac{4}{\alpha_{k}}}}\nonumber \\
\partial_{t}\alpha_{k} & = & \frac{\alpha_{k}\left(D-\frac{3}{4}K_{A}\alpha_{k}\right)}{2\pi\left(4+\tilde{\mu}_{k}^{2}-\frac{4}{\alpha_{k}}\right)}\left[\frac{2(1-\alpha_{k})+\alpha_{k}\tilde{\mu}_{k}}{1+\alpha_{k}\tilde{\mu}_{k}}
+\frac{\tilde{\mu}_{k}}{\sqrt{4+\tilde{\mu}_{k}^{2}-\frac{4}{\alpha_{k}}}}\log\frac{2+\tilde{\mu}_{k}-\sqrt{4+\tilde{\mu}_{k}^{2}-\frac{4}{\alpha_{k}}}}{2+\tilde{\mu}_{k}+\sqrt{4+\tilde{\mu}_{k}^{2}-\frac{4}{\alpha_{k}}}}\right]\,.\label{11}
\end{eqnarray}
\end{widetext}
We can recover the beta function of the dimensionless surface tension given in \cite{David_Guitter_Peliti_1987} by keeping the terms of order $\alpha\, \tilde{\mu}_k$ in (\ref{11}).
The flow is depicted in Figure 2 in the case $K_A=1$ and $D=3$ where the non--Gaussian fixed--point has values $\alpha_{*}=4$ and $\tilde{\mu}_{*}=0.107$, and is characterized by an attractive and a repulsive direction. 
Thus the inclusion of the Polyakov term in the effective average action (\ref{3.1}) had the effect of creating a non--trivial fixed--point,
giving an explicit example of how a non--local term can alter the phase portrait of a model.

The fact that non--local terms in the action modify the RG flow of the local coupling, without inducing a self--renormalization,
is reminiscent of the effect WZWN term has on the flow of the NLSM coupling \cite{Witten:1983ar}.   

Hexatic membranes have a continuous phase transition in two dimensions and are characterized by a continuous symmetry (re--parametrization invariance),
but since the phase transition is driven by long--range Coulomb interactions between Gaussian curvatures,
there is no contradiction with 
the Mermin-Wagner-Hohenberg theorem \cite{MWH1}.
%
The fixed--point value $\alpha_*$ depends continuously on the
hexatic stiffness $K_A$ and one actually have a line of fixed--points as in the Kosterlitz--Thouless phase transition \cite{Kosterlitz}.
As we will see in a moment, also the critical exponents depend continuously on $K_A$.

At the non--Gaussian fixed point the membrane has a non--trivial fractal dimension.
This is related to the mass critical exponent by the general relation $d_{F}=\frac{1}{\nu}$ \cite{Nelson_Piran_Weinberg_1988}, where $\nu$ is minus the inverse of the negative eigenvalue of the stability matrix. 
Linearizing the flow around the non--Gaussian fixed--point gives the estimate:
\begin{eqnarray}
d_{F} &=&  2+\frac{D-2}{\pi}\frac{D}{3K_{A}}\nonumber \\
&&+\frac{D-2}{2}\,\tilde{\mu}_{*}\left(\frac{D}{3K_{A}}\right)^{\frac{3}{2}}+\frac{2(D-2)}{\pi}\,\tilde{\mu}_{*}^{2}\left(\frac{D}{3K_{A}}\right)^{2}\nonumber\\
&&+\frac{3(D-2)}{4}\,\tilde{\mu}_{*}\left(4+\tilde{\mu}_{*}^{2}\right)\left(\frac{D}{3K_{A}}\right)^{\frac{5}{2}}+...\label{df}
\end{eqnarray}
where $\tilde{\mu}_{*}$ is the fixed point value of the dimensionless
surface tension. The first correction term in (\ref{df}) agrees
with the one found in \cite{David_Guitter_Peliti_1987}. The fractal
dimension is shown in Figure 3 as a function of the hexatic stiffness
for $D=3$; since its value is bigger than the classical dimension $d=2$,
the membrane is crinkled, i.e. it is infinitely rugged but still spatially extended.
Our estimate (\ref{df}) is strictly lower than the leading $1/K_A$ result for all values
of the hexatic stiffness and it tends to $d_{F}=2.709$ as $K_{A}\rightarrow 0$.
\begin{figure}
\begin{centering}
\includegraphics[scale=0.65]{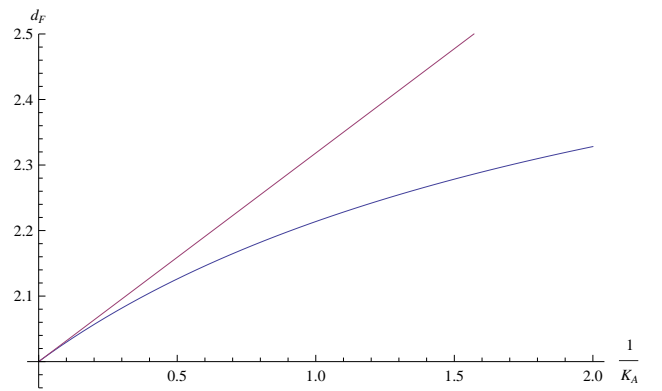}
\par\end{centering}
\caption{The fractal dimension $d_{F}$ as a function of $1/K_A$ for $D=3$.
The upper curve is the leading order term in $1/K_A$ while the lower curve is our estimate (\ref{df}).}
\end{figure}

\subsection{Conclusions}

In this paper we studied hexatic membranes using a geometrical approach based on the effective average action.
The $O(2)$--model coupled to the fluid membrane has a non--trivial RG fixed--point where the membrane
is crinkled and has fractal dimension depending on the hexatic stiffness and on the embedding dimension with values between two and three.

This example shows how matter coupled to fluctuating geometries can induce physically interesting phases through the generation of non--local invariants.
The calculation of the present paper gives an example of the non--trivial role played by non--local terms in the effective action,
showing that their influence can drastically change the phase structure of a given model.
In this light, it is important to consider the influence of non--local terms in other models of fluctuating geometry,
such as membranes in higher dimensions \cite{Codello:2012dx} or  quantum gravity  \cite{Codello:2011js}.

This work also opens the road to the study of matter--membranes systems by the methods of the effective average action and the relative exact flow equation.
The question that naturally arises is how flat space universality classes of $O(N)$--models \cite{Codello:2012ec},
other than $N=2$, are dressed by the geometrical fluctuations of the membrane.
Are there infinitely many fixed--points in the $N=0,1$ cases? Does the Mermin-Wagner-Hohenberg theorem applies to the $N\geq2$ cases?
This, and the inclusion of topological excitations in the $N=2$ case, will be subject of further studies.

\subsubsection*{Acknowledgments}

We would like to thank D.~Mouhanna for stimulating discussions and the LPTMC for hospitality.
The research of O.Z. is supported by the DFG within the Emmy-Noether program (Grant SA/1975 1-1). 

\appendix

\section{
}
We report here some details about the calculation of the contributions
to the beta functions of $\kappa_{k}$ and $\bar{\kappa}_{k}$ induced
by the Polyakov action:
\begin{equation}
I_{P}[g]=\int\sqrt{g}\, R\frac{1}{\Delta}R\,.\label{A_1}
\end{equation}
Since we are expanding the rhs of the flow equation (\ref{2}) to
order $K^{2}$ or $R$, it is enough to keep only contributions of
order $R$ in the Hessian of the Polyakov action (\ref{A_1}). The
expansion is conveniently written as:
\begin{eqnarray}
\delta^{2}I_{P}[g]=\left.2\int d^{2}x\sqrt{g}\,\delta R\,\frac{1}{\Delta}\,\delta R\;\right|_{\delta g_{\mu\nu}=h_{\mu\nu}}\qquad\qquad\nonumber\\
+\left.2\int d^{2}x\sqrt{g}\,R\,\frac{1}{\Delta}\,\delta R\;\right|_{\delta g_{\mu\nu}=H_{\mu\nu}}+O(R^{2},RK^2)\,,\quad\label{A_2}
\end{eqnarray}
with $h_{\mu\nu}=-2\nu^{i}K_{\alpha\beta}^{i}$ the first variation
of the metric and $H_{\mu\nu}=2(\partial_{\alpha}\nu^{i}\partial_{\beta}\nu^{i}+\nu^{i}\nu^{j}K_{\alpha}^{i\gamma}K_{\gamma\beta}^{j})$
the second. After the expansion is performed, we can set the background metric, and implicitly the embedding field, equal to the metric of the two dimensional sphere; in this way we make all derivatives of the curvatures vanish, but we will still be able to disentangle the operators $R$ and $K^2$. Using the variation $\delta R=g^{\mu\nu}\Delta\delta g_{\mu\nu}+\nabla^{\mu}\nabla^{\nu}\delta g_{\mu\nu}-\frac{1}{2}g^{\mu\nu}\delta g_{\mu\nu}R$
we find:
\begin{widetext}
\begin{eqnarray}
\int d^{2}x\sqrt{g}\,R\frac{1}{\Delta}\delta R
&=&\int d^{2}x\sqrt{g}\,H_{\mu\nu}\left(Rg^{\mu\nu}+\nabla^{\mu}\nabla^{\nu}\frac{1}{\Delta}R\right)+O(R^{2},RK^2)\nonumber\\
&=&\int d^{2}x\sqrt{g}\frac{R}{2}\,\nu^{i}\Delta\nu^{i}+O(R^{2},RK^2)\nonumber\\
\int d^{2}x\sqrt{g}\,\delta R\frac{1}{\Delta}\delta R 
&=& \int d^{2}x\sqrt{g}\left(h\Delta h+2h\nabla^{\mu}\nabla^{\nu}h_{\mu\nu}+h_{\alpha\beta}\frac{1}{\Delta}\nabla^{\alpha}\nabla^{\beta}\nabla^{\mu}\nabla^{\nu}h_{\mu\nu}\right)+O(R^{2})\nonumber \\
 & = & \int d^{2}x\sqrt{g}\left(\frac{1}{2}K^{i}K^{j}+K_{\alpha\beta}^{i}K^{j\alpha\beta}\right)\nu^{i}\Delta\nu^{j}+O(R^{2})\,.\label{A_4}
\end{eqnarray}
\end{widetext}
In (\ref{A_4}) we made the substitutions:
\begin{eqnarray*}
\nabla^{\mu}\nabla^{\nu} & \rightarrow & \frac{1}{2}\nabla^{2}g^{\mu\nu}\\
\nabla^{\alpha}\nabla^{\beta}\nabla^{\mu}\nabla^{\nu} & \rightarrow & \frac{1}{8}\nabla^{4}\left(g^{\alpha\beta}g^{\mu\nu}+g^{\alpha\mu}g^{\beta\nu}+g^{\alpha\nu}g^{\beta\mu}\right)\,,
\end{eqnarray*}
that are allowed, to this order, under the trace in the flow equation and we used the relation
$K_{\alpha\beta}K^{\alpha\beta}=K^{2}-R$ to simplify the first term. Inserting 
(\ref{A_4}) in (\ref{A_2}) gives the contribution (\ref{3.2}) of the Polyakov
action (\ref{A_1}) to the Hessian (\ref{3.01}) of the effective average action.

\end{document}